\pdfoutput=1

\newcommand*{\ATLASLATEXPATH}{latex/}
\documentclass[cernpreprint, texlive=2016, UKenglish]{\ATLASLATEXPATH atlasdoc}

\usepackage{\ATLASLATEXPATH atlaspackage}
\usepackage{\ATLASLATEXPATH atlasbiblatex}
\usepackage{\ATLASLATEXPATH atlasphysics}

\graphicspath{{logos/}{figures/}}
\usepackage{HIGG-2016-20-PAPER-defs}

\usepackage{comment}
\usepackage{color}
\usepackage{feynmp-auto}
\usepackage{multirow}             
\usepackage{siunitx}              
\usepackage{latex/atlascover}

\AtlasTitle{Search for Higgs boson pair production in the $\gamma\gamma WW^*$ channel using \pp collision data recorded at \sqs$= $\SI{13}{\TeV} with the ATLAS detector}

\author{The ATLAS Collaboration}

\date{\today}

\PreprintIdNumber{CERN-EP-2018-104}

\AtlasRefCode{HIGG-2016-20}

\AtlasDate{\today}

\AtlasJournal{EPJC}

\AtlasJournalRef{Eur. Phys. J. C 78 (2018) 1007} 
\AtlasDOI{10.1140/epjc/s10052-018-6457-x}

\AtlasAbstract{
  Searches for non-resonant and resonant Higgs boson pair production are performed
  in the $\gamma\gamma WW^*$ channel with the final state of $\gamma\gamma\ell\nu jj$
  using \SI{36.1}{\ifb} of proton--proton collision data
  recorded at a centre-of-mass energy of $\sqrt{s}=$\SI{13}{\TeV} by the ATLAS detector at the Large Hadron Collider.
  No significant deviation from the Standard Model prediction is observed.
  A 95\% confidence-level observed upper limit of \SI{7.7}{\pico\barn}
  is set on the \xs for non-resonant production, while the expected limit is \SI{5.4}{\pico\barn}.
  A search for a narrow-width resonance $X$ decaying to a pair of Standard Model Higgs bosons $HH$ 
  is performed with the same set of data,
  and the observed upper limits on $\sigma(pp\rightarrow X)\times B(X\rightarrow HH)$
  range between \SI{40.0}{\pico\barn} and \SI{6.1}{\pico\barn}
  for masses of the resonance between \SI{260}{\GeV} 
  and \SI{500}{\GeV}, while the expected limits range between \SI{17.6}{\pico\barn} and \SI{4.4}{\pico\barn}.
  When deriving the limits above, the Standard Model branching ratios of the \Hyy and \HWW are assumed. 
}

\hypersetup{pdftitle={ATLAS draft},pdfauthor={The ATLAS Collaboration}}

\begin{document}

\maketitle

\section{Introduction}
\label{sec:intro}

A particle consistent with the Standard Model (SM) Higgs boson ($H$) was discovered by both the ATLAS and CMS experiments
at the Large Hadron Collider (LHC) in 2012~\cite{HIGG-2012-27,CMS-HIG-12-028}.
Various studies of its properties have been performed~\cite{HIGG-2013-17,HIGG-2014-06,CMS-HIG-14-009,CMS-HIG-14-018,CMS-HIG-15-002},
and no significant deviation from the SM predictions has been found.
The SM Higgs boson is a strong probe of physics beyond the SM.
This paper documents searches for both non-resonant and resonant production of Higgs boson pairs (\hh)
in the semileptonic \wwyy final state using \SI{36.1}{\ifb} of proton--proton
(\pp) collision data recorded by the ATLAS detector at a centre-of-mass energy of $\sqs = \SI{13}{\TeV}$.
Previous searches for Higgs boson pair production 
have been performed by both the ATLAS and CMS experiments 
with data recorded at $\sqs = \SI{8}{\TeV}$ in the final states 
\bbbb~\cite{Aad:2015uka}, \bbyy~\cite{HIGG-2013-29,Khachatryan:2016sey}, \bbtautau~\cite{Aad2015xja,Khachatryan:2015tha,Sirunyan:2017tqo}
and \wwyy~\cite{Aad2015xja}, as well as multi-lepton and multi-photon~\cite{Khachatryan:2014jya}.
The \pp collision data at $\sqs = \SI{13}{\TeV}$
have been analysed in order to search for Higgs boson pairs in the final states
\bbbb~\cite{Aaboud:2018knk} and \bbww~\cite{Sirunyan:2017guj}.
No significant excess was observed compared to the SM prediction.
However, it is important to explore the $\SI{13}{\TeV}$ data in the channels that are not covered yet,
such as the \wwyy channel presented in this paper. Although this decay channel is not the most sensitive 
amongst all possible Higgs boson decays, it relies on the Higgs boson couplings to vector bosons, which 
are already relatively well measured. Furthermore, this channel will contribute to the final combination 
of all measurable $HH$ decays.

\begin{figure}[!h!tbp]
  \centering
  \subfloat[]{\includegraphics[width=0.315\textwidth]{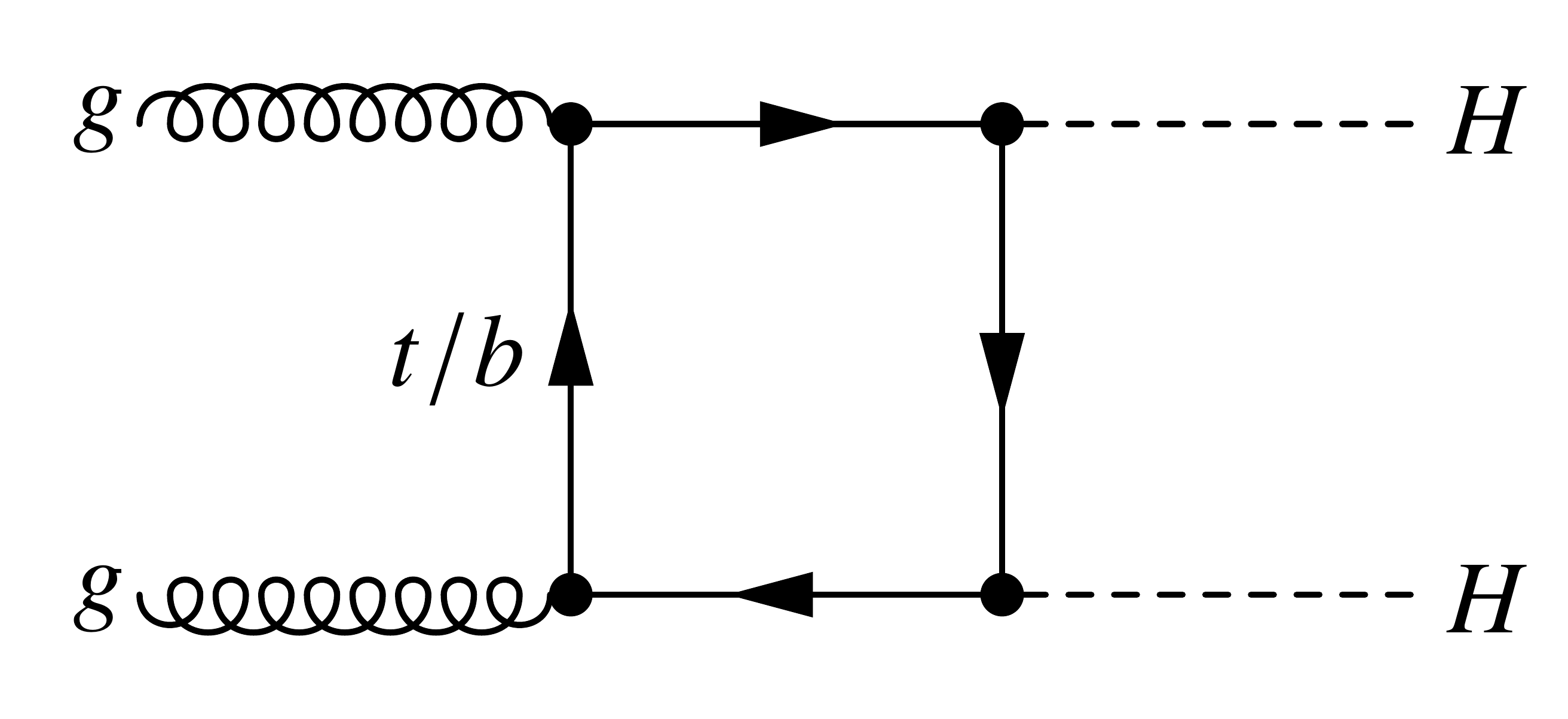}}
  \subfloat[]{\includegraphics[width=0.315\textwidth]{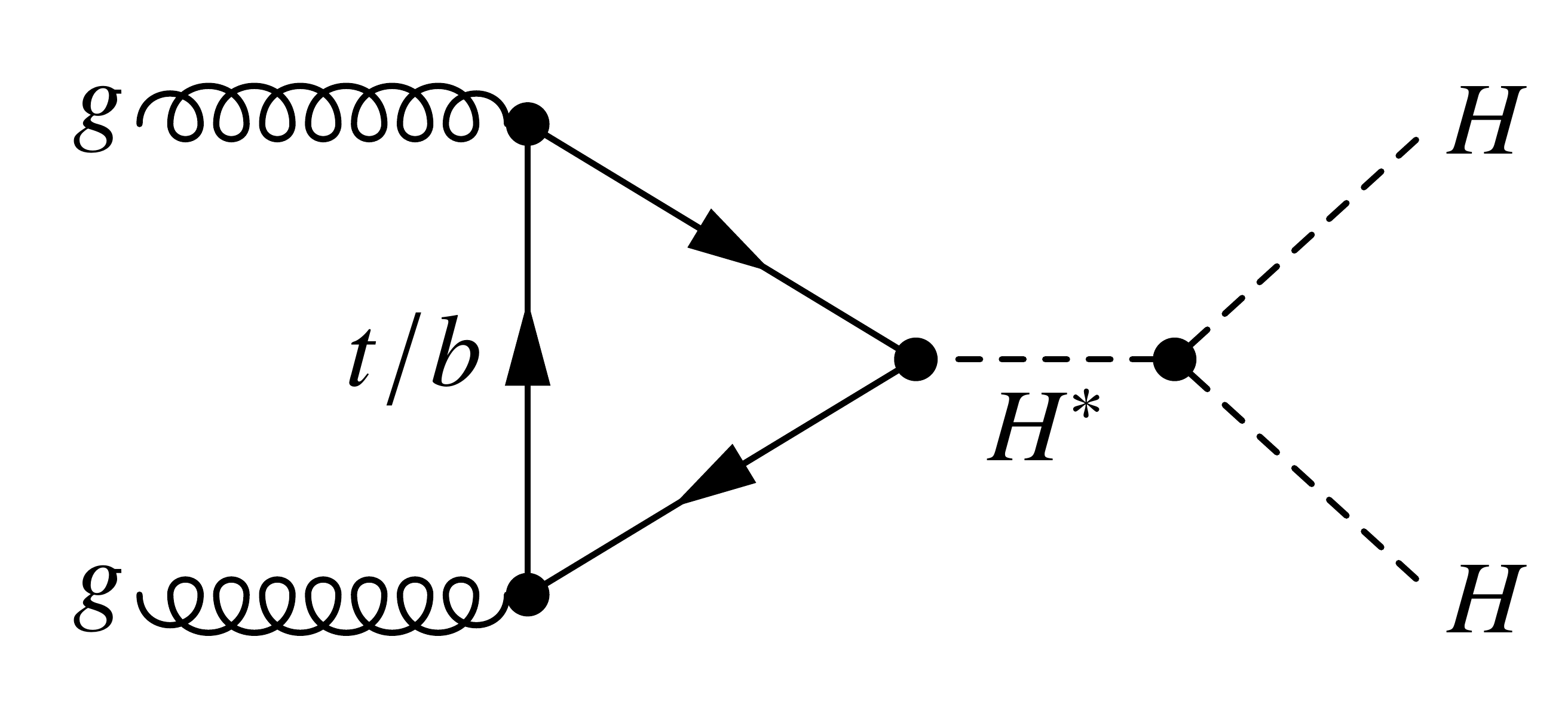}}
  \subfloat[]{\includegraphics[width=0.315\textwidth]{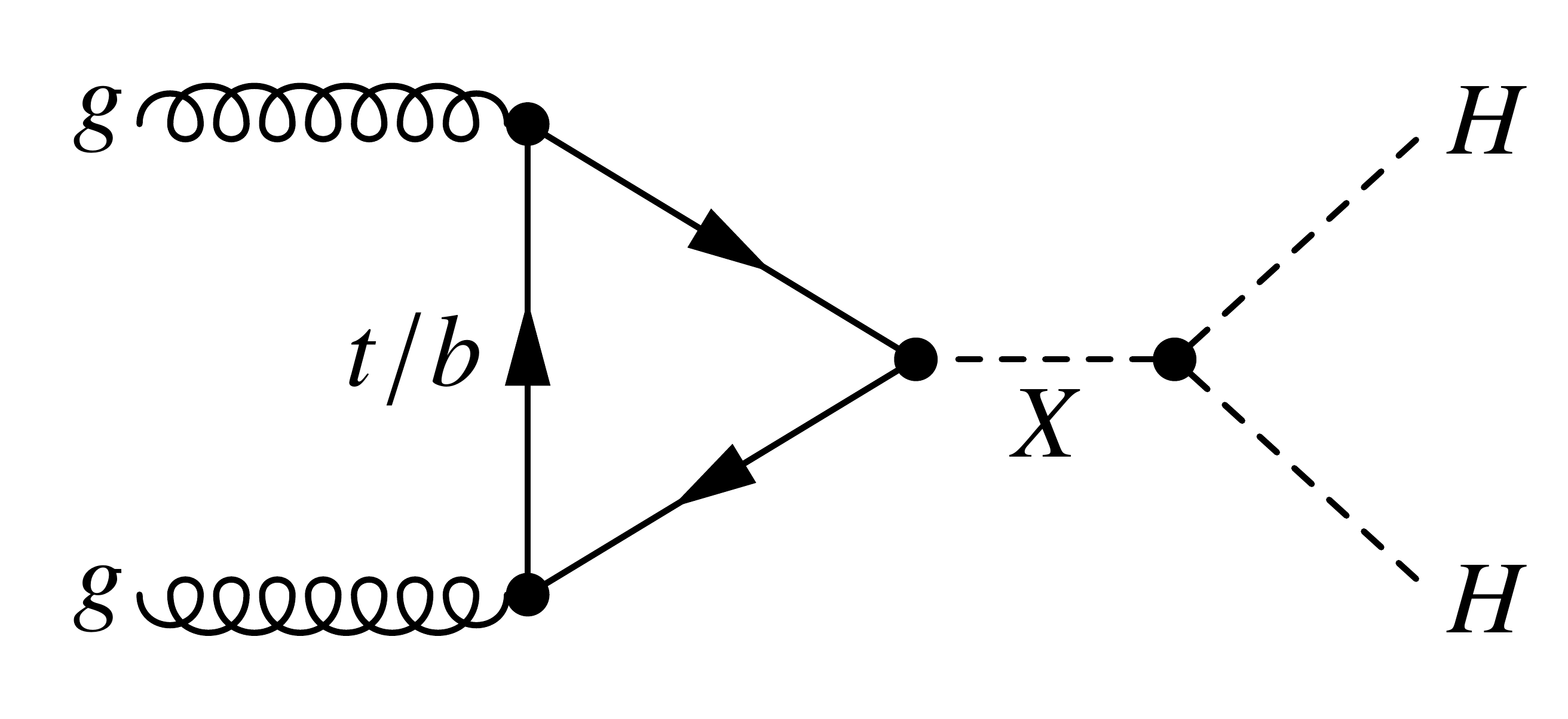}}
  \caption{
    Feynman diagrams for leading-order Higgs boson pair production in the SM through
    (a) a heavy-quark loop, (b) the Higgs self-coupling, and (c) an intermediate heavy resonance in a BSM scenario.
    The total SM contribution is the sum of the two modes depicted in (a) and (b), which have 
    significant destructive interference.
    Physics beyond the SM can enhance Higgs boson pair production
    either by modifying the Higgs boson coupling constants from their SM values in (a) and/or (b),
    or by an additional s-channel exchange of an intermediate scalar resonance in (c).}
  \label{fig:production_diagrams}
\end{figure}

The SM prediction of the Higgs boson pair production \xs is several orders
of magnitude smaller than the single-Higgs-boson production rate~\cite{yr4},
due to additional $ttH$ or $HHH$ vertices, an additional on-shell Higgs boson that reduces the kinematic phase space,
and the fact that the two leading-order (LO) Feynman diagrams have
strong destructive interference~\cite{Glover:1987nx}.
In Figure~\ref{fig:production_diagrams}(a),
the so-called box diagram represents Higgs boson pair production via a heavy-quark loop,
where the \xs scales with the squared value of the $ttH$ or $bbH$ coupling constants.
In Figure~\ref{fig:production_diagrams}(b),
the so-called triangle diagram contributes to Higgs boson 
pair production via the exchange of a virtual Higgs boson
and is the only tree-level diagram sensitive to the Higgs boson self-coupling constant ($\lambda_{HHH}$),
the squared value of which scales the \xs.

In many beyond-the-SM (BSM) scenarios, Higgs boson pair production can be enhanced
by modifying the $ttH$, $bbH$ or $\lambda_{HHH}$
coupling constants from their SM values, reducing the effect of
the destructive interference~\cite{Baglio:2012np} between the diagrams of Figures~\ref{fig:production_diagrams}(a) and (b),
or by replacing the virtual Higgs boson with an intermediate scalar resonance, cf. Figure~\ref{fig:production_diagrams}(c).
Various BSM models with extended Higgs sectors predict a heavy Higgs boson 
decaying into a pair of Higgs bosons similar to the one in the SM. 
Such models include the two-Higgs-doublet models (2HDM)~\cite{Branco:2011iw},
the minimal supersymmetric extension of the SM~\cite{Dimopoulos:1981zb},
twin Higgs models~\cite{Chacko:2005vw} and composite Higgs models~\cite{Grober:2010yv,Mrazek:2011iu}.
Heavy resonances, other than heavy Higgs bosons, that can decay into a pair of SM Higgs bosons, are predicted in different models,
 and could for instance be gravitons~\cite{Tang:2012pv}, radions~\cite{Cheung:2000rw} or stoponium~\cite{Kumar:2014bca}.

This paper reports searches for non-resonant and resonant production of pairs of Higgs bosons
in the semileptonic \wwyy final state ($\gamma\gamma\ell\nu jj$), i.e. with 
two photons, two jets, one charged lepton and a neutrino.
This final state benefits from the large branching fraction of \HWW~\cite{yr4},
a characteristic signature from two photons and one lepton,
as well as the excellent resolution of the \diphoton invariant mass \myy,
which provides good discrimination from a smooth continuum background composed of multi-photon and multi-jet SM processes.
Given the expected sensitivity in $\SI{13}{\TeV}$ data,
the di-Higgs-boson mass range between 260~GeV and 500~GeV is explored 
in the search for a scalar resonant Higgs boson pair production.

\section{The ATLAS detector}
\label{sec:atlas}

\newcommand{\AtlasCoordFootnote}{
ATLAS uses a right-handed coordinate system with its origin at the nominal
interaction point (IP) in the centre of the detector and the $z$-axis along the beam pipe.
The $x$-axis points from the IP to the centre of the LHC ring, and the $y$-axis points upwards.
Cylindrical coordinates $(r,\phi)$ are used in the transverse plane,
$\phi$ being the azimuthal angle around the $z$-axis.
The pseudorapidity is defined in terms of the polar angle $\theta$ as $\eta = -\ln \tan(\theta/2)$.
Angular distance is measured in units of $\Delta R \equiv \sqrt{{(\Delta\eta)}^{2} + {(\Delta\phi)}^{2}}$.}

The ATLAS experiment~\cite{PERF-2007-01} is a multipurpose particle detector
with a forward-backward symmetric cylindrical geometry and nearly
$4\pi$ coverage in solid angle.\footnote{\AtlasCoordFootnote}
It consists of an inner tracking detector (ID) surrounded by
a thin superconducting solenoid providing a \SI{2}{\tesla} axial magnetic field,
electromagnetic (EM) and hadronic calorimeters, and a muon spectrometer (MS).
The ID covers the pseudorapidity range $\abseta < 2.5$
and consists of silicon pixel, silicon microstrip, and transition-radiation tracking systems.
The innermost pixel layer, the insertable B-layer~\cite{Capeans:2010jnh},
was installed at a mean radius of 3.3~cm after Run 1,
and has been operational since the beginning of Run 2.
Lead/liquid-argon (LAr) EM sampling calorimeters with high granularity provide energy measurements of EM showers.
A steel/scintillator-tile hadronic calorimeter covers the central pseudorapidity range ($\abseta < 1.7$).
The endcap and forward regions are covered by LAr calorimeters for
EM and hadronic energy measurements up to $\abseta = 4.9$.
The MS surrounds the calorimeters and is based on three large
air-core toroid superconducting magnets with eight coils each
and with bending power in the range \SIrange[range-units = single]{2.0}{7.5}{\tesla\metre}.
It includes a system of fast detectors for triggering purposes and precision tracking chambers.
A dedicated two-level trigger system is used to select events~\cite{Aaboud:2016leb}.
The first-level trigger is implemented in hardware and
uses a subset of the detector information to reduce the accepted event rate to at most \SI{100}{\kilo\hertz}.
This is followed by a software-based trigger level that
reduces the accepted event rate to an average of \SI{1}{\kilo\hertz}.

\section{Data and simulated samples}
\label{sec:datamc}

\subsection{Data samples}

The full set of \pp collision data collected during 2015 and 2016 
are used in this analysis. The two datasets were recorded at
the same centre-of-mass energy \sqs$=$\SI{13}{\TeV}, albeit with different beam conditions.
Beam intensities in 2016 were typically higher than in 2015,
resulting in a higher instantaneous luminosity and a larger number of
\pp collisions in each bunch crossing.
The integrated luminosity of the combined 2015+2016 dataset used in this analysis is $36.1\pm\SI{0.8}{\per\femto\barn}$.
This dataset were collected in run periods during which all subsystems were operational.
The events are collected with a trigger requiring the presence of at least two photons,
one with a transverse energy \ET $>$ 35~GeV and the second with \ET $>$ 25~GeV,
and the longitudinal and transverse profiles of the EM shower were 
required to be consistent with those expected for a photon.
The corresponding trigger efficiency reaches about 99\%
for the events that pass the event selection of the analysis.

\subsection{Simulated event samples}
\label{sec:simulevtsample}

Simulated Monte Carlo (MC) samples are used to estimate the signal acceptance and study the modelling
for both non-resonant SM Higgs boson pair production and resonant BSM Higgs boson pair production.
MC samples are also used to estimate the acceptance and study the modelling for SM single-Higgs-boson production
processes, and to study the modelling of the SM continuum background from events
with multiple photons and jets (Section~\ref{sec:modelling}), which is the dominant background in the analysis.
Eventually, it is estimated by a data-driven method for both its normalisation and shape. 

\begin{table*}[htbp]
\begin{center}
\caption{Simulated signal samples.}
\label{tab:MCSig}
\begin{tabular}{lllll}
\toprule
 Processes                   & Generator  &Parton shower  &Tune & PDF \\
\hline
 Non-resonant                & \MGaMCatNLO 2.2.3   &\textsc{Herwig}++       &UEEE5 &CTEQ6L1 \\
 Resonant                    & \MGaMCatNLO 2.2.3   &\textsc{Herwig}++       &UEEE5 &CTEQ6L1 \\
\bottomrule
\end{tabular}
\end{center}
\end{table*}

The simulated samples for signals are listed in Table~\ref{tab:MCSig}.
The event generator \MGaMCatNLO 2.2.3~\cite{Alwall:2014hca} was used for the production of
non-resonant~\cite{Frederix:2014hta} and resonant~\cite{Hespel:2014sla} signal MC samples 
at next-to-leading order (NLO) in QCD, where four values of the resonance 
mass ($m_{X}$ = 260, 300, 400 and 500~GeV) are considered. 
The events were generated by a Higgs Effective Field Theory (HEFT) 
using the \MCatNLO method~\cite{Frixione:2002ik}
and were reweighted in order to take into account the effects of the finite top-quark mass.
The parton shower was implemented using \textsc{Herwig}++ 2.7.1~\cite{Bellm:2013hwb} with a set of 
tuned underlying-event parameters called the UEEE5 tune\footnote{The tune parameters can be found at the following 
link: \url{https://herwig.hepforge.org/tutorials/mpi/tunes.html}}~\cite{Gieseke:2012ft}, 
and the parton distribution function (PDF) set CTEQ6L1~\cite{Nadolsky:2008zw} was used.
\begin{itemize}
 \item For non-resonant Higgs boson pair production, the inclusive \xs{s} are normalised
to the SM prediction of \SI{33.41}{\femto\barn}~\cite{Grazzini:2018bsd, yr4}, calculated at NNLO in QCD,
including resummation of soft-gluon emission at next-to-next-to-leading-logarithmic (NNLL) accuracy,
as prescribed by the LHC Higgs \XS Working Group~\cite{yr4}.
The effect of the finite top-quark mass is also taken into account at NLO~\cite{Borowka:2016ehy}.
 \item For resonant Higgs boson pair production, a narrow decay width,
which is negligible compared to the experimental mass resolution, is assumed.
The interference between non-resonant
and resonant Higgs boson pair production is implemented in the generator.
The interference is minimal and remains negligible when a narrow decay width is assumed.
\end{itemize}

\begin{table*}[htbp]
\begin{center}
\setlength\tabcolsep{5pt}
\caption{Simulated SM single-Higgs-boson background samples with $m_{H} =$~125~GeV.}
\label{tab:singleH}
\resizebox{0.95\textwidth}{!}{\begin{tabular}{lllllll}
\toprule
Processes   & Generators                & QCD order & EW order & PDF          &Parton shower  &Normalisation\\
\hline
ggF         & \textsc{Powheg NNLOPS}    & NNLO    & NLO & \textsc{PDF4LHC15}  &\PYTHIA 8.186  & N$^3$LO (QCD) + NLO (EW) \\
VBF         & \textsc{Powheg}           & NLO     & NLO & \textsc{PDF4LHC15}  &\PYTHIA 8.186  & NNLO (QCD) + NLO (EW) \\
$W^{+}H$    & \textsc{Powheg MiNLO}     & NLO     & NLO & \textsc{PDF4LHC15}  &\PYTHIA 8.186  & NNLO (QCD) + NLO (EW) \\
$W^{-}H$    & \textsc{Powheg MiNLO}     & NLO     & NLO & \textsc{PDF4LHC15}  &\PYTHIA 8.186  & NNLO (QCD) + NLO (EW) \\
$q\bar{q}\rightarrow ZH$          & \textsc{Powheg MiNLO}     & NLO     & NLO & \textsc{PDF4LHC15}  &\PYTHIA 8.186  & NNLO (QCD) + NLO (EW) \\
$ggZH$      & \textsc{Powheg MiNLO}     & NLO     & NLO & \textsc{PDF4LHC15}  &\PYTHIA 8.186  & NLO NLL (QCD)  \\
$t\bar{t}H$ & \textsc{MadGraph aMC@NLO} & NLO     & NLO & \textsc{NNPDF3.0}   &\PYTHIA 8.186  & NLO (QCD) + NLO (EW) \\
\bottomrule
\end{tabular}
}
\end{center}
\end{table*}

Table~\ref{tab:singleH} lists the simulated samples for
the dominant SM single-Higgs-boson production modes:
gluon--gluon fusion (\ggH), vector-boson fusion (\VBFH), associated production with a $W$ or $Z$ boson (\VH),
and associated production with a pair of top quarks (\ttH).
For all these processes, the \PYTHIA 8.186 parton shower is used
for the modelling of non-perturbative effects. 
The AZNLO tune~\cite{STDM-2012-23} is used in \ggH, \VBFH and \VH simulations,
while the A14 tune is used in \ttH simulations.
\begin{itemize}
 \item \textbf{gluon--gluon fusion:} The \ggH production is accurate to NNLO in QCD, using 
the \POWHEG method~\cite{Bagnaschi:2011tu} for matching the matrix element with the parton shower, 
and the MiNLO method~\cite{Hamilton:2012np,Hamilton:2012rf} to 
simultaneously achieve NLO accuracy for inclusive Higgs boson production. Furthermore,
a reweighting procedure was performed using the HNNLO program~\cite{Catani:2007vq,Grazzini:2008tf,Grazzini:2013mca}
to achieve full NNLO accuracy~\cite{Hamilton:2013fea}. This sample is referred to as NNLOPS.
The PDF4LHC15 NLO PDF set~\cite{Butterworth:2015oua} was used.
The inclusive \xs of the \ggH production is normalised to the calculation at next-to-next-to-next-to-leading-order (N$^3$LO)
QCD and NLO electroweak (EW) accuracies~\cite{Anastasiou:2016cez}.
 \item \textbf{\VBFH and \VH:} \VBFH and \VH production was simulated at NLO in QCD with \POWHEGBOX
v2~\cite{Bagnaschi:2011tu,Nason:2009ai,Alioli:2010xd} using the PDF4LHC15 NLO PDF set.
The inclusive \VBFH contribution is normalised to the \xs calculated with NLO QCD and NLO EW
corrections~\cite{Ciccolini:2007ec,Ciccolini:2007jr,Arnold:2008rz} with 
an approximate NNLO QCD correction applied~\cite{Bolzoni:2010xr}.
The contributions are normalised to \xs{s} calculated with NNLO QCD~\cite{Brein:2003wg}
and NLO EW corrections~\cite{Ciccolini:2003jy} for $WH$ and $q\bar{q}\rightarrow ZH$ and 
at NLO and next-to-leading-logarithm (NLL) accuracy in QCD for $gg\rightarrow ZH$~\cite{Altenkamp:2012sx}.
 \item \textbf{\ttH:} The \ttH production is simulated using \MGaMCatNLO 2.2.3 and
its inclusive \xs is normalised to a calculation with NLO QCD and NLO EW
corrections~\cite{Beenakker:2001rj,Beenakker:2002nc,Dawson:2002tg,Dawson:2003zu}.
\end{itemize}

Processes of continuum backgrounds of multiple photons and jets with either one
or zero leptons were simulated with \MGaMCatNLO 2.2.2, interfaced with the 
parton shower model in \PYTHIA 8.186.

Multiple \pp collisions in each bunch crossing, ``pile-up'',
were simulated with the soft QCD processes of \PYTHIA 8.186 using the A2
tune~\cite{ATL-PHYS-PUB-2012-003} and the MSTW2008LO PDF set~\cite{Martin:2009iq}.
An additional event-level reweighting is performed in order to ensure that the distribution
of the average number of interactions per bunch crossing matches that occurring in the data used in this analysis.
The particles in the final states of the generated processes were passed through
either a \geant~\cite{Agostinelli:2002hh} simulation of the ATLAS detector,
or through the ATLAS fast simulation framework~\cite{SOFT-2010-01}, which has been extensively validated against the \geant simulation model.
The output from the detector simulation is then analysed using the same reconstruction software as the data.
The MC samples for single-Higgs-boson production were simulated with the \geant framework,
while the other samples used in this analysis were produced with the ATLAS fast simulation framework.

\section{Object and event selection}
\label{sec:selection}

The event selection is based on the properties of the visible objects in the final state,
which includes one charged lepton (electron or muon), two jets, and two photons.
These objects are reconstructed from detector-level objects, such as
energy clusters in the EM calorimeter and tracks in the ID, as well as in the MS.

\subsection{Object reconstruction}
\label{subsec:objectselection}

Photon candidates are reconstructed from clusters of energy deposited in the EM calorimeter~\cite{ATL-PHYS-PUB-2016-015}.
If the candidates are matched with a reconstructed conversion vertex or tracks consistent
with the hypothesis of a \yee conversion, they are classified as converted photon candidates.\footnote{Converted photons
are those that convert to an $e^{+}e^{-}$ pair inside 
the ID volume, with at least one of the two lepton trajectories 
reconstructed, while unconverted photons directly enter the EM calorimeter.}
If the matched track is consistent with the hypothesis of an electron produced 
in the beam interaction region, they are classified as electron candidates.
If the candidates are not matched with a reconstructed conversion vertex or tracks satisfying the conversion requirement
they are classified as unconverted photon candidates.
The energy is determined by summing the energies of all cells
that belong to the associated cluster~\cite{Aaboud:2016yuq}
and is corrected using a combination of simulation-based and
data-driven calibration factors~\cite{PERF-2013-05}
determined from $Z\rightarrow e^+e^-$ events collected in 2015 and 2016. The
photon energy resolution in simulation is corrected to match the resolution in data~\cite{ATL-PHYS-PUB-2016-015}.
The reconstructed photon candidates are required to meet ``tight'' photon identification
criteria~\cite{Aaboud:2016yuq}, which are based on
the lateral and longitudinal energy profiles of EM showers in the calorimeter.
The identification efficiency is measured as a function of the transverse energy of photons (\ETy).
It ranges from 90\% to 98\% for converted photons
and from 85\% to 95\% for unconverted photons,
in the \ETy interval between 25 and 200 \GeV.
To suppress the background from jets misidentified as photons,
all reconstructed photon candidates are required to meet a set of
calorimeter- and track-based isolation criteria~\cite{STDM-2010-08}.
A calorimeter-based isolation variable $E^{\text{iso}}_{\text{T}}$ is 
defined as the sum of the transverse energies (\ET) of
all positive-energy topological clusters of calorimeter cells~\cite{Aad:2016upy} 
within $\Delta R = 0.2$ of the photon candidate,
excluding the energy of the photon candidate itself.
The selection applied to the calorimeter-based isolation variable is $E^{\text{iso}}_{\text{T}} < 0.065\ET^{\gamma}$.
A track-based isolation variable $p^{\text{iso}}_{\text{T}}$ is defined as the scalar 
sum of the transverse momenta (\pT) of tracks
with \pT $>$ \SI{1}{\GeV} within $\Delta R = 0.2$ of the photon candidate, excluding tracks from photon conversions.
The selection on the track-based isolation variable is $p^{\text{iso}}_{\text{T}} < 0.05E_{\text{T}}^{\gamma}$.
Only photon candidates with \abseta $<$ 2.37 are considered,
excluding the transition region between the barrel and endcap calorimeters (1.37 $<$ \abseta $<$ 1.52).

Electron candidates are reconstructed from clusters of energy deposited in the EM calorimeter
matched to a track in the inner detector, as described above.
A likelihood-based (LH) algorithm is used~\cite{ATLAS-CONF-2016-024}
to perform the electron identification against the background from jets or non-prompt
electrons. Electron candidates are identified according to the ``medium LH'' criteria.
Muon candidates are identified by matching a reconstructed ID track with
a reconstructed MS track~\cite{PERF-2015-10}.
The identification classifies muon candidates as either ``loose'' or ``medium'',
based on the number of hits in the different
ID and MS subsystems, and on the significance of the difference $|q/p_{\text{MS}}-q/p_{\text{ID}}|$,
where $q$ is the charge and $p$ is the momentum of the muon candidate,
as well as on the energy deposit in the tile hadronic calorimeters.
The ``medium'' candidates are used in the analysis.
An efficiency ranging from 84\% to 93\% as a function of \ET or \pT is achieved 
in the combined identification and reconstruction of electrons,
and 96\% (above 98\%) in muon identification (reconstruction),
in the range where the objects are selected.
The electron (muon) is required to pass the ``Loose'' (``GradientLoose'') isolation criterion 
based on the sum of \pT of tracks lying within a cone of 
$\Delta R=$ min (10 GeV/$p_{\text{T}}^{e (\mu)}$, 0.3) 
and the sum of \ET of topological clusters of calorimeter cells 
within a cone of $\Delta R=$ 0.2 (0.2) around the 
electron (muon) candidate, excluding the contributions from the electron (muon) candidate. 
With these requirements the isolation efficiencies for electrons (muons) are above
99\% (0.057$p_{\text{T}}^{\mu}+$95.57\%)~\cite{ATLAS-CONF-2016-024, PERF-2015-10}. 
Finally, the electron candidates are required to have $\ET > 10$~GeV and $|\eta|<2.47$, excluding
the transition region between the barrel and endcap calorimeters (1.37 $<$ \abseta $<$ 1.52),
whereas the muon candidates are required to have $p_{\text{T}} > 10$~GeV and $|\eta|<2.7$.

Jets are reconstructed via the FastJet package~\cite{Cacciari:2011ma}
using the \antikt clustering algorithm~\cite{Cacciari:2008gp}
with a radius parameter $R=$ 0.4.
The jet energies are determined at the EM scale and calibrated using particle-level correction factors
based on a combination of simulation and data~\cite{PERF-2011-03,PERF-2012-01,PERF-2011-04,PERF-2014-03,ATL-PHYS-PUB-2015-015}.
Jets are required to have \abseta $<$ 2.5 and \pT $>$ \SI{25}{\GeV}.
In addition, a jet-vertex tagging algorithm (JVT) ~\cite{ATL-PHYS-PUB-2014-001} is applied to
jets with \abseta $<$ 2.4 and \pT $<$ \SI{60}{\GeV} in order to suppress jets originating from pile-up interactions.
In this algorithm, a multivariate discriminant based on two track-based variables is
constructed to reject pile-up jets while maintaining a high efficiency for the hard-scatter jet
independent of the number of primary vertices in the event.
The selected jets are classified as \bjet{s} 
using a multivariate technique~\cite{PERF-2012-04,ATL-PHYS-PUB-2015-022},
which takes advantage of the information about secondary vertices,
the impact parameters of the associated tracks and the topologies 
of decays of heavy-flavour hadrons.
The $b$-tagging working point is selected to have an
efficiency of 70\% for a $b$-jet from $t\bar{t}$ decays, with
a rejection factor of 12 for jets originating from $c$-quarks ($c$-jets),
and of close to 400 for jets initiated by light-flavour quarks or gluons (light-flavour jets).

An overlap removal procedure is performed in the following order to avoid double counting of detector-level objects
when reconstructing physics objects.
Electrons with $\Delta R(e,\gamma)<$ 0.4 are removed.
Jets with $\Delta R(\text{jet},\gamma)<$ 0.4 or $\Delta R(\text{jet},e)<$ 0.2 are removed.
Electrons with $\Delta R(e,\text{jet})<$ 0.4 are removed.
Muons with $\Delta R(\mu,\gamma)<$ 0.4 or $\Delta R(\mu,\text{jet})<$ 0.4 are removed.

\subsection{Event selection}
\label{subsec:eventselection}

\begin{figure}[!h!tpb]
  \centering
  \includegraphics[width=0.49\textwidth]{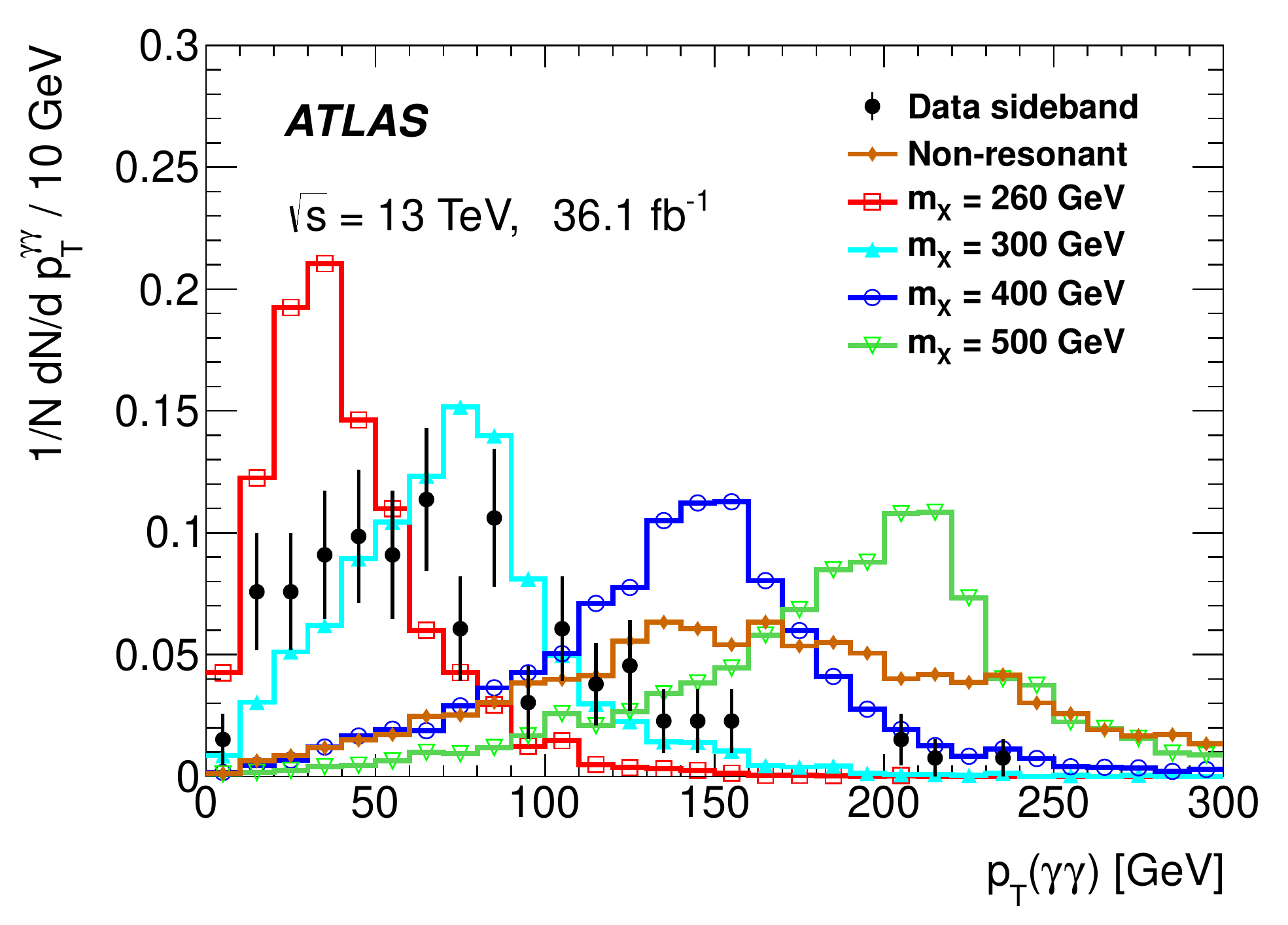}
  \caption{Distributions of the reconstructed transverse momenta of the \diphoton system with all event 
           selections, except the \pTyy selection, applied for various 
           signal models, as well as sideband data, normalised to unit area.}
  \label{fig:ptyy}
\end{figure}

The events passing the \diphoton trigger are required
to contain at least two jets, no \bjet, and at least one charged lepton ($e$ or $\mu$, but including contributions from 
fully leptonic $\tau$-lepton decays) in the final state.
The two photon candidates with the leading (sub-leading) \ET are required to
satisfy $\ET^{\gamma}/\myy >$ 0.35 (0.25).
The \bjet veto suppresses the \ttH process.
Furthermore, the transverse momentum of the diphoton system (\pTyy)
is required to be larger than 100~GeV for maximising the sensitivity 
and keeping at least 70\% of signal events. This requirement suppresses continuum background events
when searching for non-resonant Higgs boson pair production,
or resonant production with resonance masses of 400~GeV or higher.
However, the \pTyy selection is omitted in the search for resonance masses below 400~GeV
due to a limited separation between signal and continuum background in this kinematical region,
as can be seen in Figure~\ref{fig:ptyy}. These final selection criteria,
together with a requirement on the invariant \diphoton mass of 105 GeV < \myy < 160 GeV,
define the event sample on which the signal search is performed for the various assumed signal models.
A data ``sideband'' sample is selected applying the same criteria, but excluding the Higgs mass region \myy 121.7--128.5 GeV,
and can be used together with other samples to study the continuum background.

If there were an observable signal, one of the Higgs bosons would be directly visible in the \myy distribution.
The combination of two jets and at least one charged lepton would be consistent with \HWW 
for the other Higgs boson. 
Its signature would in principle be enhanced by a missing transverse energy (\met) 
requirement to indicate a neutrino, but a selection on \met was found not to 
produce any significant improvement in sensitivity, and so was not applied. 
The magnitude of \met~\cite{ATL-PHYS-PUB-2015-027, ATL-PHYS-PUB-2015-023} 
is measured from the negative vectorial sum of the transverse momenta of all
photon, electron and muon candidates and of all hadronic jets after accounting for 
overlaps between jets, photons, electrons, and muons, as well as an estimate of
soft contributions based on tracks.

After all selections described above, the combined acceptance and selection efficiency for non-resonant production
is 8.5\%, while it ranges from 6.1\%
to 10\% as a function of the mass of the resonance ($m_X$) from 260~GeV to 500~GeV,
as shown in Table~\ref{tab:acceptance}.
The efficiency for the non-resonant Higgs boson pair production is at the same level as the efficiency for the high-mass resonant production,
as the Higgs bosons and their decay products tend to exhibit large transverse momenta due 
to the box diagram shown in Figure~\ref{fig:production_diagrams}(a).

\FloatBarrier
\begin{table*}[htbp]
\begin{center}
  \caption{ The combined acceptance and efficiency for non-resonant and resonant 
            with different scalar resonance masses $m_{X}$, with and without a \pTyy selection.}
  \label{tab:acceptance}
  \begin{tabular}{ lccccccc }

  \toprule
                         & \multicolumn{3}{c}{No \pTyy selection} & \phantom{X} & \multicolumn{3}{c}{\pTyy $>$ 100~GeV} \\
  \hline
    $m_{X}$ [GeV]        & 260 & 300 & 400 & \phantom{X} & 400 & 500 & Non-resonant \\
  \hline
    Acceptance $\times$ efficiency [\%]      & 6.1 &7.1  & 9.7 & \phantom{X} &7.8  &10  &8.5 \\
  \bottomrule
  \end{tabular}
\end{center}
\end{table*}

\section{Signal and background estimation}
\label{sec:modelling}

A fit to the \myy distribution is performed to extract the signal yield
as described in Section~\ref{sec:result}.
The shapes of both the signal and background distributions are modelled with analytical functions.
For both Higgs boson pair production and single-Higgs-boson processes, 
the \myy distributions are modelled with 
double-sided Crystal Ball functions~\cite{Oreglia:1980cs}.
Their shape parameters are determined by a fit to simulated samples.
The single-Higgs-boson contribution is normalised to the SM cross-sections as described in Section~\ref{sec:simulevtsample}.
Higgs boson pair production is regarded as a background to the resonant search.
Its contribution is also set to the SM prediction of Section~\ref{sec:simulevtsample}.

The continuum background is modelled with an exponential function of a second-order polynomial.
Several functional forms were evaluated by fitting the sidebands 
in data and MC samples under different conditions of photon purity and lepton multiplicity.
Photon purity was lowered, compared to the final data selection, by 
reversing the requirements on photon isolation or identification.
For higher photon purity, MC samples with prompt photons were used. 
The lepton multiplicity was varied to be zero or at least one.
For all combinations of conditions, the exponential function with a second-order polynomial
gave the best fits, with satisfactory $\chi^2$, and was chosen to model the continuum background. 
The shape parameters and normalisation are free to float in the final fit to the data. 
Since any functional form might introduce spurious signals, this effect 
is estimated with a sample mixing irreducible prompt-photon background from
simulation and reducible backgrounds from data,
as described in Section~\ref{sec:uncertainties}.

The expected numbers of signal and background events are shown in Table~\ref{tab:eventyields}
together with the number of events observed in data. Only events 
within a mass window of $\mH \pm 2$~\sigmamyy are reported, 
where the Higgs boson mass (\mH) is taken to be 125.09 GeV~\cite{Aad:2015zhl} and
the \diphoton mass resolution (\sigmamyy) is 1.7 GeV and is obtained from simulation. The dominant background 
is from continuum processes with multiple photons and jets.
A small background arises from SM single-Higgs-boson production processes, among which
\ttH and $WH$ productions give the leading contributions with, respectively, a 
fraction of 41.5\% (39.2\%) and 23.3\% (22.5\%) of the whole single-Higgs-boson
contribution with (without) the \pTyy $>$ 100~GeV selection.

\FloatBarrier
\begin{table*}[htbp]
\begin{center}
  \caption{
    Numbers of expected and observed events in the $\mH \pm 2$~\sigmamyy mass window with or without a \pTyy selection.
    A \xs of \SI{33.41}{\femto\barn} is assumed for non-resonant Higgs boson pair production
    when it is considered as a background in resonant searches.
    The resulting yields are determined from the fit to data by integrating the resulting functional forms
    over the selected \myy range.
    The error in each yield includes both the statistical and 
    systematic uncertainties, as discussed in Section~\ref{sec:uncertainties}.
  }\label{tab:eventyields}

  \begin{tabular}{ l r @{\,\( \pm \)\,} l  r @{\,\( \pm \)\,} l }
  \toprule
    Process              & \multicolumn{4}{c}{Number of events} \\
	                 & \multicolumn{2}{c}{No \pTyy selection} & \multicolumn{2}{c}{\pTyy $>$ 100~GeV} \\
  \midrule
    Continuum background & 22     & 5                             & 5.1    & 2.3 \\
    SM single-Higgs      & 1.92   & 0.15                          & 1.0    & 0.09\\
    SM \diHiggs          & 0.046  & 0.004                         & 0.038   & 0.004 \\

  \midrule
    Sum of expected background & 24     & 5                       & 6.1    & 2.3 \\
  \midrule
    Data                 & \multicolumn{2}{l}{\hspace{3.9mm} 33}  & \multicolumn{2}{l}{\hspace{5.2mm} 7} \\
  \bottomrule
  \end{tabular}
\end{center}
\end{table*}

\FloatBarrier
\section{Systematic uncertainties}
\label{sec:uncertainties}

\subsection{Theoretical uncertainties}

Theoretical uncertainties in the prediction of the cross section 
of single Higgs bosons are estimated from variations of the 
normalisation and factorisation scales, PDF,
and the running QCD coupling constant ($\alpha_\text{S}$) \cite{yr4}.
Among the dominant production modes \ttH and \VH,
the \xs of \ttH has the largest uncertainty: up to 9.2\% in the scale variations,
up to 3.0\% in the PDF variations, and 2.0\% in the $\alpha_\text{S}$ variations,
as prescribed by the LHC Higgs \XS Working Group~\cite{yr4}.

The theoretical uncertainties in the efficiency times acceptance ($\epsilon\times A$)
are estimated from scale, PDF and parton shower variations.
The scale uncertainty ranges from 2.1\% to 4.1\% for resonant production 
and is 3.4\% for non-resonant production.
The PDF uncertainty is around 2.5\% and 3.0\% for the resonant and non-resonant production, respectively.
The parton shower uncertainty is estimated by comparing \PYTHIA 8 and Herwig++ as two different shower models,
and ranges from 6.0\% at $m_{X} = 500$~GeV to 29.6\% at $m_{X} = 260$~GeV for resonant production,
and is 7.8\% for non-resonant production. This uncertainty is large in low-mass resonant production
because the jet spectrum at low-$p_\text{T}$ is more susceptible to variations in the parton shower model.
Non-resonant Higgs boson pair production is considered as a background
in the search for resonant Higgs boson pair production.
The scale, PDF, $\alpha_\text{S}$ and HEFT uncertainties
in the calculation of the \xs for SM Higgs boson pair production are also taken into account.
These values are 6.0\%, 2.1\%, 2.3\%, and 5.0\%, respectively, following the recommendations in Ref.~\cite{yr4}.
Further uncertainties arising from the \Hyy and \HWW branching ratios ($B$) are considered as well.
They are 2.1\% and 1.5\%~\cite{yr4}, respectively.

\subsection{Modelling uncertainties in the continuum background}

The exponential function of a second-order polynomial is determined
to provide the simplest and most robust functional form for modelling
the continuum background
as described in Section~\ref{sec:modelling}.
The uncertainties in the modelling are estimated
by fitting a signal-plus-background model to a simulated background-only sample
that has such a large number of events that its own statistical uncertainty does not affect the test results.
The fitted number of signal events ($n_\text{ss}$) quantifies spurious signal events.
The fits are performed with the assumed $m_H$ ranging from 120 GeV to 130 GeV in
steps of 0.5 GeV. The maximum value of the fitted signal yields |$n_\text{ss}$| is
regarded as a bias in the yields due to the background modelling (the spurious signal), and is, conservatively,
taken into account in the fit as the modelling uncertainty.
The fitted |$n_\text{ss}$| value reaches as large as 0.46 when not applying the $\pTyy$ selection,
and 0.26 when applying the selection.
The simulated background-only samples include the irreducible process of $\gamma\gamma\ell\nu jj$
and the reducible processes represented by events where one or two hadronic jets are misidentified as photons.
The reducible processes are modelled by the data events with reversed photon identification
or isolation requirements. The two components are combined according to the measured diphoton purity,
which is about 88\% (90\% with \pTyy selection) and normalised according to 
the number of selected data events.

\subsection{Experimental uncertainties}

The uncertainty in the measurement of the combined 2015+2016 integrated luminosity is 2.1\%. It is 
derived, following a methodology similar to that detailed in Ref.~\cite{Aaboud:2016hhf}, from a 
calibration of the luminosity scale using $x$--$y$ beam-separation scans performed in August 2015 and May 2016.
All processes that are estimated using simulation are affected by the uncertainty in the luminosity measurement.

The efficiency of the \diphoton trigger is estimated using bootstrap
methods~\cite{PERF-2011-02} with a systematic uncertainty of 0.4\%.
The photon identification uncertainty is obtained by varying the data-to-simulation efficiency corrections
within their uncertainties, derived from control samples of photons from radiative $Z$ boson
decays and from inclusive $\gamma$ events, and of electrons from $Z\rightarrow e^+e^-$ decays.
A maximal uncertainty of 1.7\% in the yields is evaluated in all of the 
SM single-Higgs-boson, SM di-Higgs-boson and BSM Higgs boson production processes.
The photon--track isolation uncertainty is derived from measurements of the uncertainty in
the data-to-simulation efficiency corrections using inclusive-photon control samples,
while the uncertainty from the calorimeter isolation requirement is evaluated from the difference between
applying and not applying corrections derived from inclusive-photon events
to the calorimeter isolation variable in the simulation.
In general, the overall isolation uncertainty is less than 1\%.
The uncertainties from the photon energy resolution and scale affect the yields by less than 0.2\%.
The relevant impact on the shape of the diphoton invariant mass is also considered by
introducing variations of the resolution and mean values of the fit function and is estimated using simulation.
The photon energy resolution varies the resolution of the $m_{\gamma\gamma}$ shape by 5.2\% to 11.4\%,
while the photon energy scale affects the mean value by about 0.5\%.
The jet energy scale (JES) and the corresponding uncertainties are 
derived from both simulation and in situ calibration using data~\cite{PERF-2012-01,Aaboud:2017jcu}.
This affects the event selection efficiency by 2.4\% to 9.9\%, depending on the process.
The jet energy resolution (JER) uncertainty is evaluated by smearing jet 
energies according to the systematic uncertainties of the resolution
measurement~\cite{ATL-PHYS-PUB-2015-015,ATL-PHYS-PUB-2015-036},
and its impact on the event selection efficiency ranges from 0.1\% to 1.6\%.
The $b$-tagging uncertainties is derived separately for $b$-jets, $c$-jets and light-flavour jets~\cite{PERF-2012-04}.
Overall, their impact on the yields is not more than 4\%.
Uncertainties arising from the reconstruction, identification and isolation of both the electron and muon candidates~\cite{ATLAS-CONF-2016-024,PERF-2015-10},
are propagated to the event yield variations, and they are found to have an impact of less than 1\%.
Finally, the pile-up reweighting procedure, which matches the distribution
of the number of interactions per bunch crossing between simulation and data, has associated
systematic uncertainties of less than 1\%. 
All experimental uncertainties are correlated among all processes that use simulation to model the yields and the kinematics.
A summary of the systematic uncertainties in the expected yields of the di-Higgs-boson and single-Higgs-boson production
is presented in Table~\ref{tab:systematic_uncertainties}.
In the search for non-resonant Higgs boson pair production, SM Higgs boson pair production is considered to be
the signal process, while single Higgs boson production is considered to be a background.
In the search for the resonant Higgs boson pair production, both SM single Higgs boson production
and non-resonant Higgs boson pair productions are considered to be background processes.

\FloatBarrier
\begin{table*}[htbp]
\begin{center}
  \setlength\tabcolsep{5pt}
  \caption{
    Summary of relative systematic uncertainties, in percent, propagated to the yields for the MC-estimated processes.
    Entries marked by `-' indicate that the systematic uncertainty is not applicable for the corresponding process.
    The extrapolation uncertainties in $b$-tagging include two components:
    one is from the extrapolation to high-\pT (\pT$>$\SI{300}{\GeV}) jets
    and the other one is from extrapolating $c$-jets to $\tau$-jets.
    The values for resonant production shown here assume $m_X$ = 260~GeV.
    Several theoretical uncertainties are reported for the cross section ($\sigma$)
    and the combined efficiency and acceptance ($\epsilon\times A$).
  }\label{tab:systematic_uncertainties}

  \begin{tabular}{ l l S S S S }
  \toprule
    \multicolumn{2}{l}{Source of uncertainties}        & {Non-resonant \hh}   & {\Xhh}  & {Single-$H$ bkg}  & {Single-$H$ bkg} \\
    \multicolumn{2}{l}{}                               &  &  & {\pTyy $>$ 100~GeV} & {No \pTyy selection} \\

  \midrule                                                                                                     
    \multicolumn{2}{l}{Luminosity 2015+2016}           & 2.1       & 2.1           &  2.1                 & 2.1          \\

  \midrule
    \multicolumn{2}{l}{Trigger}                        & 0.4       & 0.4           &  0.4                 & 0.4          \\
    \multicolumn{2}{l}{Event sample size}              & 1.7       & 2.2           &  1.6                 & 1.3          \\
    \multicolumn{2}{l}{Pile-up reweighting}            & 0.5       & 0.9           &  0.7                 & 0.6          \\

  \midrule
    \multirow{4}{*}{Photon}    & identification        & 1.7        & 1.4           &  0.8                & 0.8          \\
                               & isolation             & 0.8        & 0.7           &  0.4                & 0.4          \\

                               & energy resolution     & 0.1        & 0.1           &  0.2                & < 0.1      \\
                               & energy scale          & 0.2        & < 0.1         &  0.2                & < 0.1      \\

  \midrule
    \multirow{2}{*}{Jet}       & energy scale          & 4.0        & 9.9           &  2.4                & 2.6          \\
                               & energy resolution     & 0.1        & 1.6           &  0.5                & 1.0          \\
  \midrule
    \multirow{4}{*}{$b$-tagging} & $b$-hadron jets     & < 0.1      & < 0.1         &  3.8                & 3.6          \\
                               & $c$-hadron jets       & 1.5        & 1.0           &  0.7                & 0.6          \\
                               & light-flavour jets    & 0.3        & 0.3           &  0.1                & 0.1          \\
                               & extrapolation         & < 0.1      & < 0.1         &  0.1                & < 0.1      \\
  \midrule
    \multirow{2}{*}{Lepton}    & electron              & 0.5        & 0.7           &  0.2                & 0.2          \\
                               & muon                  & 0.5        & 0.7           &  0.3                & 0.5          \\

  \midrule
    \multirow{7}{*}{Theory}    & PDF on $\sigma$       & 2.1        & {-}           &  3.4                 & 3.4         \\
                               & $\alpha_S$ on $\sigma$& 2.3        & {-}           &  1.3                 & 1.3         \\
                               & scale on $\sigma$     & 6.0        & {-}           &  0.9                 & 0.9         \\
                               & HEFT on $\sigma$      & 5.0        & {-}           &  {-}                 & {-}         \\
                               & scale on $\epsilon\times A$        & 2.8           & 2.5      &  {-}      & {-}         \\
                               & PDF on $\epsilon\times A$          & 3.0           & 2.4      &  {-}      & {-}         \\
                               & parton shower on $\epsilon\times A$   & 7.8           & 29.6        &  {-}        & {-} \\
                               & $B$(\Hyy)   & 2.1           & 2.1           &  2.1        & 2.1          \\
                               & $B$(\HWW)   & 1.5           & 1.5           &  1.5        & 1.5          \\
  \midrule                                                                                          
    \multicolumn{2}{l}{Total}                          & 13.6          & 31.8       &  7.1      &  6.8    \\
  \bottomrule
  \end{tabular}
\end{center}
\end{table*}

\FloatBarrier
\section{Results}
\label{sec:result}

A fit to $m_{\gamma\gamma}$ is performed in the signal region to extract the signal yield.
The statistical model is constructed with a likelihood function:
\begin{align}
  \begin{split}
  \mathcal{L}(\mu,\thetaBF) = 
     & \prod_{i} \Big(( n_{\mathrm{Signal}}(\mu,\thetaBF) + n_{\mathrm{ss}} ) \times f^{1}_{\mathrm{DSCB}}\Big(m_{\gamma\gamma}^i,\thetaBF\Big) + n_{\mathrm{Cont}} \times f_{\mathrm{Cont}}\Big(m_{\gamma\gamma}^i,\thetaBF\Big) +\\ 
     & n_{\text{SM-one-Higgs}}(\thetaBF) \times f^{2}_{\mathrm{DSCB}}\Big(m_{\gamma\gamma}^i,\thetaBF\Big) +\\
     & n_{\text{SM-di-Higgs}} \times f^{3}_{\mathrm{DSCB}}\Big(m_{\gamma\gamma}^i,\thetaBF\Big) \Big) \prod G(0|\thetaBF,1)
  \end{split}
\label{eq:lr}
\end{align}

\begin{itemize}
\item $i$ stands for the event index,
\item $n_{\mathrm{Signal}}$ is the expected number of signal events,
\item $\mu$ is the cross section (times the branching fraction of $X\rightarrow HH$) of non-resonant (resonant) production,
\item $n_{\mathrm{ss}}$ is the estimated spurious signal yield due to our choice of continuum background modelling,
\item $f^{1}_{\mathrm{DSCB}}$ is the probability density function (pdf) of a double-sided Crystal Ball distribution for signal,
\item $n_{\mathrm{Cont}}$ is the expected number of continuum background events,
\item $f_{\mathrm{Cont}}$ is the pdf of the continuum background, i.e. an exponential function of a second-order polynomial,
\item $n_{\text{SM-one-Higgs}}$ is the expected number of single-Higgs-boson events, which is set to the SM prediction and can vary with uncertainties,
\item $f^{2}_{\mathrm{DSCB}}$ is the pdf of a double-sided Crystal Ball distribution for the SM single-Higgs-boson background,
\item $n_{\text{SM-di-Higgs}}$ is the expected number of the SM di-Higgs-boson events, 
\item $f^{3}_{\mathrm{DSCB}}$ is the pdf of a double-sided Crystal Ball distribution for SM di-Higgs-boson background,
\item $G(0|\thetaBF,1)$ is the pdf of a Gaussian distribution used to constrain the nuisance parameters $\thetaBF$ that 
model systematic uncertainties as introduced in Section~\ref{sec:uncertainties}.
\end{itemize}

Equation (\ref{eq:lr}) is used directly for the BSM resonant signal searches.
For the non-resonant SM Higgs boson pair search, the SM Higgs 
boson pair term is removed.

\FloatBarrier
\begin{figure}[tbhp]
  \centering
  \subfloat[]{\includegraphics[width=0.49\textwidth]{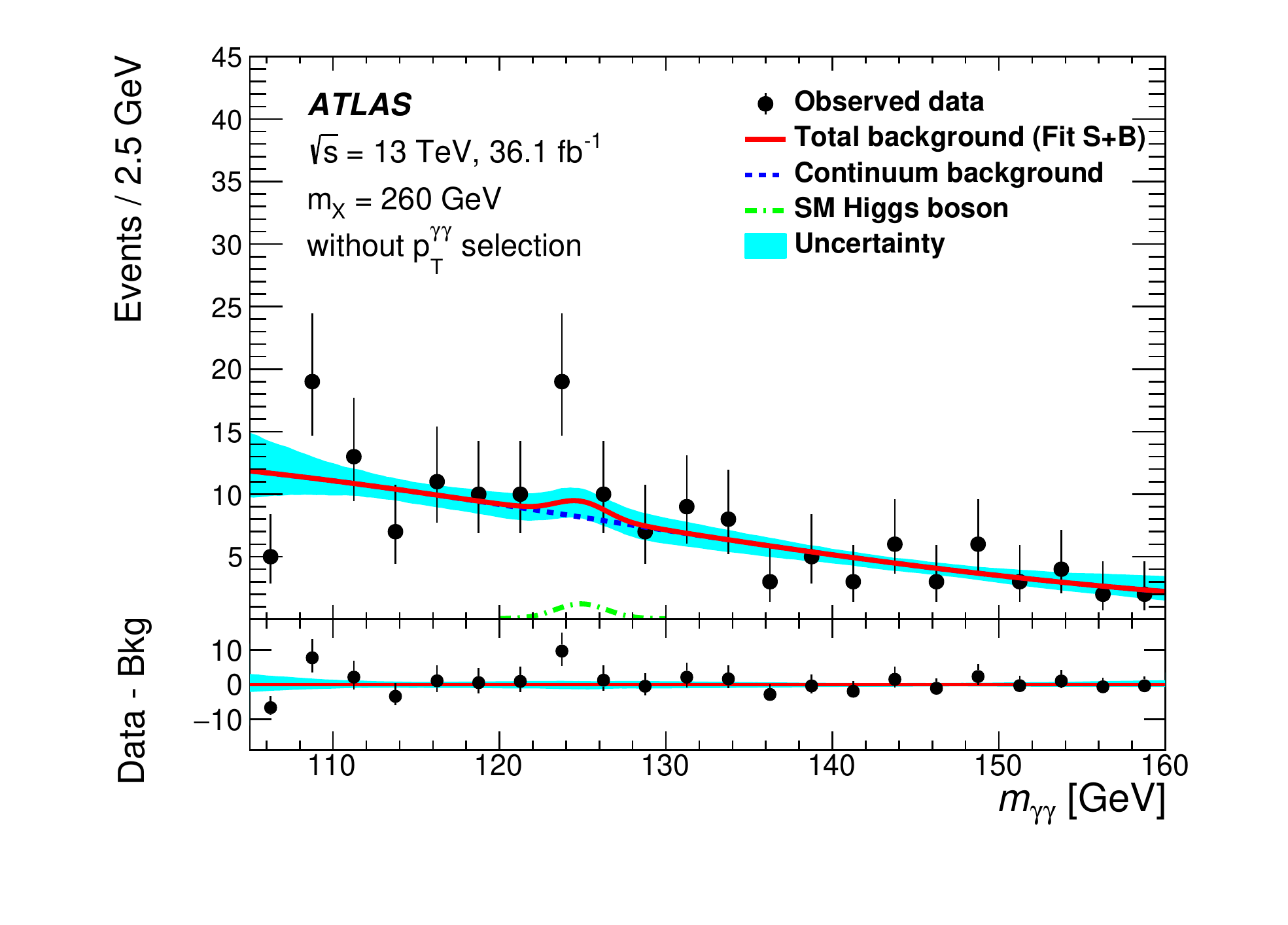}} 
  \subfloat[]{\includegraphics[width=0.49\textwidth]{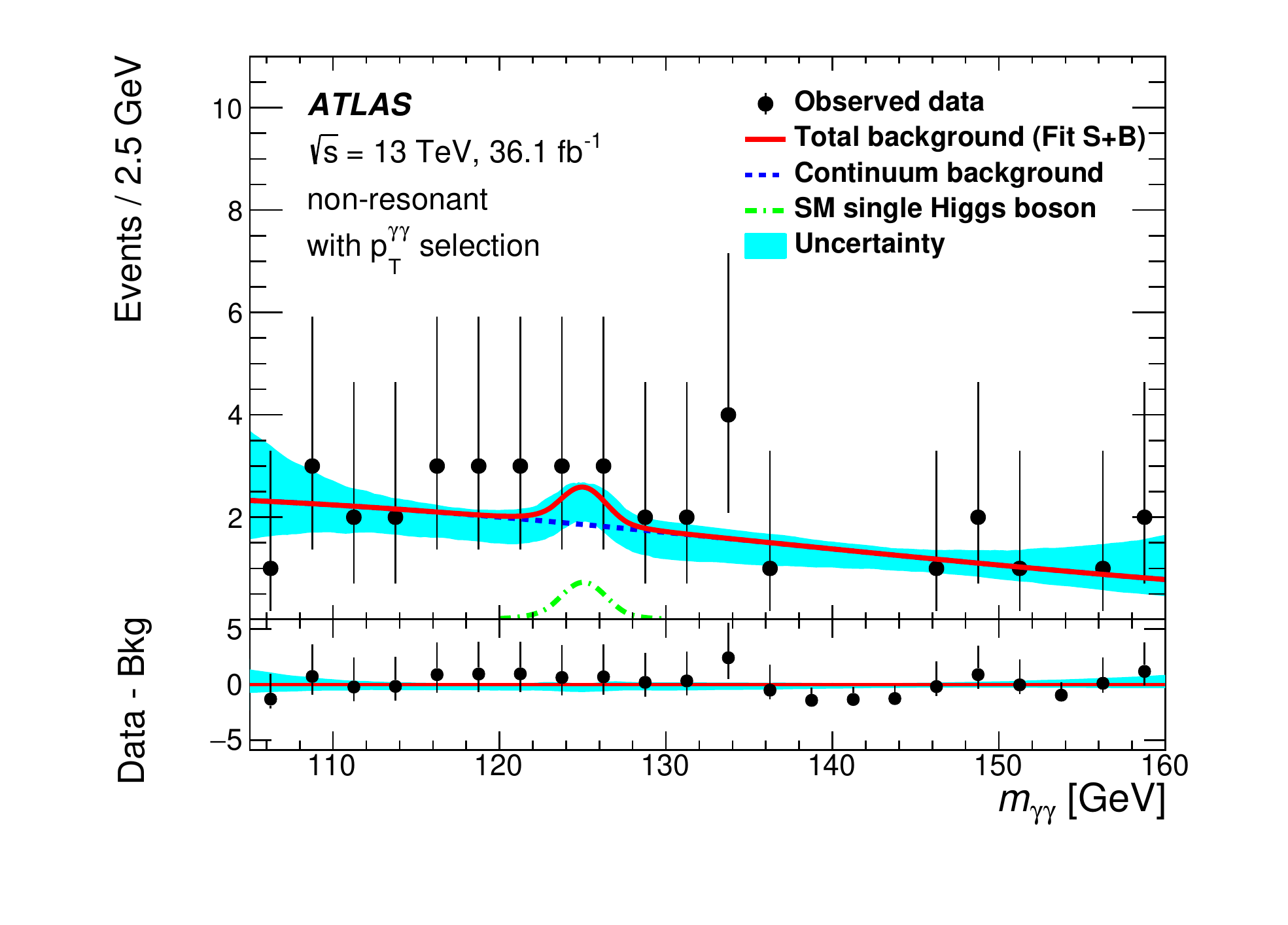}} 
  \caption{
           Invariant mass spectrum of the diphoton system in the searches
           for both resonant and non-resonant Higgs boson pair production,
           with the corresponding backgrounds for (a) $m_{X} = 260$ GeV
           without any \pTyy selection and (b) the non-resonant case with a \pTyy > 100 GeV selection.
           Fits to \myy are performed using the full signal-plus-background model.
           In each plot, only the background component is present.
           The shape parameters and normalisation of the continuum background model
           are determined in the fits.
           The ``SM Higgs boson'' in (a) contains the single-Higgs-boson background
           and SM di-Higgs-boson background.
           The band shows the uncertainty of the ``Total background'' in the upper
           panel and is calculated by a sampling method.
           The bottom panel shows the difference between the number of
           events in data and the estimated number of background events,
           as determined by the fits.
  }\label{fig:nonresfit}
\end{figure}

\begin{figure}[!htbp]
  \centering
  \subfloat[]{\includegraphics[width=0.49\textwidth]{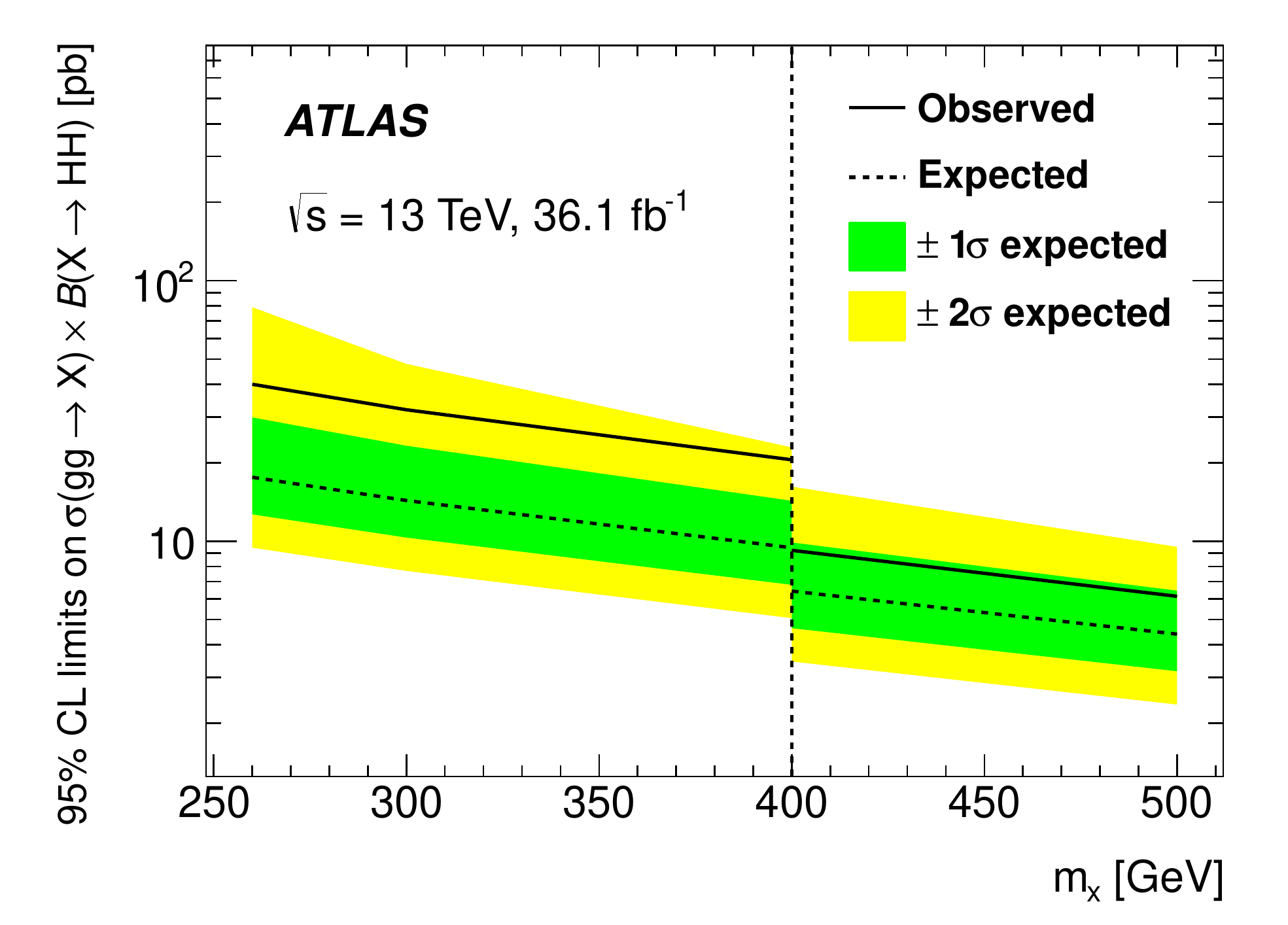}}
  \subfloat[]{\includegraphics[width=0.49\textwidth]{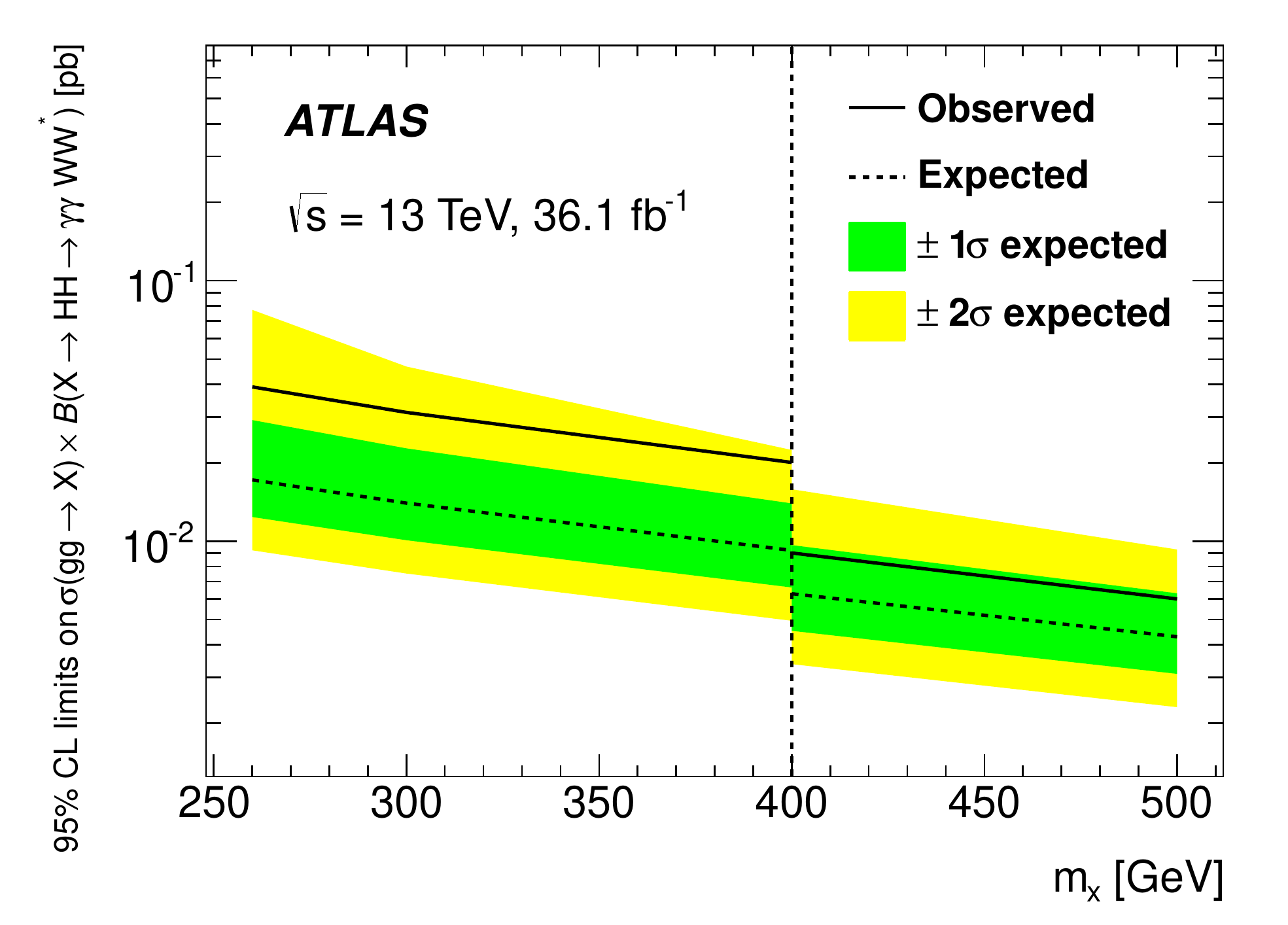}} 
  \caption{
	$95$\% CL expected (dashed line) and observed (solid line) limits on the resonant Higgs boson pair production
	\xs times the branching fraction of \Xhh as a function of \mX (a) with 
	and (b) without assuming the SM branching fractions of \HWW and \Hyy. 
        The $\text{CL}_\text{s}$ method and the asymptotic approximation are used.
        The $\pm1\sigma$ and $\pm2\sigma$ bands on the expected limit are also presented. To the right of the
        vertical dashed line at $m_{X} = 400$~GeV, the $\pTyy > 100$~GeV selection is applied in both plots, but not 
        to the left.}
  \label{fig:reslimit}
\end{figure}

The distributions in the final signal-plus-background fit using the likelihood function above
are shown for two sets of selections separately:
in Figure~\ref{fig:nonresfit}(a) without requiring the \pTyy selection for masses below 400~GeV,
and in Figure~\ref{fig:nonresfit}(b) requiring \pTyy $>$ 100~GeV for masses above 400~GeV,
as well as for the search for non-resonant Higgs boson pair production.
The fits are performed separately on the two distributions to search for resonant signals
in both the low-mass and high-mass ranges.
The observed data are found to be compatible with the sum of the expected SM backgrounds
by performing a likelihood-ratio test~\cite{Cowan:2010js}.
The largest data excess has a local significance
of 2.0 standard deviations at 400 GeV without the \pTyy selection.
A modified frequentist method
$\text{CL}_\text{s}$~\cite{Read:2002hq} is used to calculate the $95$\% confidence-level (CL) exclusion limits
with the asymptotic approximation~\cite{Cowan:2010js}.
Unfolding the SM Higgs boson branching fractions to $WW^*$ and $\gamma\gamma$ for the signal,
the expected upper limit on the \xs for non-resonant Higgs boson pair production is
\SI[parse-numbers = false]{5.4}{\pico\barn},
while the observed limit is \SI{7.7}{\pico\barn}, as shown in Table~\ref{tab:limit-nonres-ratio}.
The difference between the expected and observed limits is due to a slight excess of events in data.
The expected upper limit on the \xs times the branching fraction of $X\rightarrow HH$ ranges from
\SI[parse-numbers = false]{17.6}{\pico\barn} to \SI[parse-numbers = false]{4.4}{\pico\barn},
while the observed limit ranges from \SI[parse-numbers = false]{40}{\pico\barn} to
\SI[parse-numbers = false]{6.1}{\pico\barn}, as a function of $m_X$ between 260 and 500~GeV,
as shown in Figure~\ref{fig:reslimit}(a).

\FloatBarrier
\begin{table*}[!h!tbp]
\begin{center}
\caption{The 95\% CL upper limits for the non-resonant production and the ratios of the 
         limits to the SM cross-section value of $\sigma(pp\rightarrow HH)=33.4^{+2.4}_{-2.8}$~fb~\cite{yr4}. 
         The $\pm1\sigma$ and $\pm2\sigma$ 
         intervals around the median limit are also presented.}
\label{tab:limit-nonres-ratio}
  \begin{tabular}{lcccccc}
  \toprule
                  & $+2\sigma$ & $+1\sigma$ & Median & $-1\sigma$ & $-2\sigma$ & Observed\\
  \midrule
  Upper limits on $\sigma(HH)$ [pb] & 12 & 8.0 & 5.4 & 3.9 & 2.9   & 7.7 \\
  \midrule
  Upper limits on $\sigma(HH)\times B(\gamma\gamma WW^{*})$ [fb] & 12 & 7.8 & 5.3 & 3.8 & 2.8 & 7.5 \\
  \midrule
  Ratios of limits over the SM $\sigma(HH)$  &360 &240 &160 &120 &87 &230\\
  \bottomrule
  \end{tabular}
\end{center}
\end{table*}

Assuming the SM Higgs branching fractions of $B(\HWW)=(21.52\pm0.32)\%$ 
and $B(\Hyy)=(0.227\pm0.005)\%$~\cite{yr4},
the expected upper limit on the \xs for non-resonant production
of $HH\rightarrow \gamma\gamma WW^*$ is \SI[parse-numbers = false]{5.3}{\femto\barn},
while the observed limit is \SI{7.5}{\femto\barn}, as shown in Table~\ref{tab:limit-nonres-ratio}.
The expected upper limit on the \xs for resonant production of $X\rightarrow HH\rightarrow \gamma\gamma WW^*$ ranges from
\SI[parse-numbers = false]{17.2}{\femto\barn} to \SI[parse-numbers = false]{4.3}{\femto\barn},
while the observed limit ranges from \SI[parse-numbers = false]{39.1}{\femto\barn} to
\SI[parse-numbers = false]{6.0}{\femto\barn}, as a function of $m_X$ between 260 and 500~GeV,
as shown in Figure~\ref{fig:reslimit}(b).
The statistical uncertainty dominates in the final limits, while the impact 
of systematic uncertainties on these limits is only a few percent.

\FloatBarrier

\section{Conclusion}
\label{sec:conclusions}

This paper presents searches for non-resonant and resonant Higgs boson pair production
with a semileptonic \wwyy final state using \SI{36.1}{\ifb} of \pp\ collision data collected
at \SI{13}{\TeV} with the ATLAS detector at the LHC\@.
No significant excess above the expected SM background is observed.
A 95\% confidence-level upper limit of \SI{7.7}{\pico\barn} is set on the \xs for non-resonant production,
while the expected limit is \SI{5.4}{\pico\barn},
compared to the SM Higgs boson pair production \xs of \SI{33.4}{\femto\barn}.
The observed upper limit on the resonant production \xs times the branching fraction of $X\rightarrow HH$
ranges between \SI{40}{\pico\barn} and \SI{6.1}{\pico\barn},
while the expected limit ranges between \SI{17.6}{\pico\barn} and \SI{4.4}{\pico\barn},
for a hypothetical resonance with a mass in the range of \SIrange{260}{500}{\GeV}.
When deriving the limits above, the SM branching ratios of the \Hyy and \HWW are assumed.

\section*{Acknowledgements}

We thank CERN for the very successful operation of the LHC, as well as the
support staff from our institutions without whom ATLAS could not be
operated efficiently.

We acknowledge the support of ANPCyT, Argentina; YerPhI, Armenia; ARC, Australia; BMWFW and FWF, Austria; ANAS, Azerbaijan; SSTC, Belarus; CNPq and FAPESP, Brazil; NSERC, NRC and CFI, Canada; CERN; CONICYT, Chile; CAS, MOST and NSFC, China; COLCIENCIAS, Colombia; MSMT CR, MPO CR and VSC CR, Czech Republic; DNRF and DNSRC, Denmark; IN2P3-CNRS, CEA-DRF/IRFU, France; SRNSFG, Georgia; BMBF, HGF, and MPG, Germany; GSRT, Greece; RGC, Hong Kong SAR, China; ISF and Benoziyo Center, Israel; INFN, Italy; MEXT and JSPS, Japan; CNRST, Morocco; NWO, Netherlands; RCN, Norway; MNiSW and NCN, Poland; FCT, Portugal; MNE/IFA, Romania; MES of Russia and NRC KI, Russian Federation; JINR; MESTD, Serbia; MSSR, Slovakia; ARRS and MIZ\v{S}, Slovenia; DST/NRF, South Africa; MINECO, Spain; SRC and Wallenberg Foundation, Sweden; SERI, SNSF and Cantons of Bern and Geneva, Switzerland; MOST, Taiwan; TAEK, Turkey; STFC, United Kingdom; DOE and NSF, United States of America. In addition, individual groups and members have received support from BCKDF, CANARIE, CRC and Compute Canada, Canada; COST, ERC, ERDF, Horizon 2020, and Marie Sk{\l}odowska-Curie Actions, European Union; Investissements d' Avenir Labex and Idex, ANR, France; DFG and AvH Foundation, Germany; Herakleitos, Thales and Aristeia programmes co-financed by EU-ESF and the Greek NSRF, Greece; BSF-NSF and GIF, Israel; CERCA Programme Generalitat de Catalunya, Spain; The Royal Society and Leverhulme Trust, United Kingdom. 

The crucial computing support from all WLCG partners is acknowledged gratefully, in particular from CERN, the ATLAS Tier-1 facilities at TRIUMF (Canada), NDGF (Denmark, Norway, Sweden), CC-IN2P3 (France), KIT/GridKA (Germany), INFN-CNAF (Italy), NL-T1 (Netherlands), PIC (Spain), ASGC (Taiwan), RAL (UK) and BNL (USA), the Tier-2 facilities worldwide and large non-WLCG resource providers. Major contributors of computing resources are listed in Ref.~\cite{ATL-GEN-PUB-2016-002}.

\printbibliography

\clearpage 
 
\begin{flushleft}
{\Large The ATLAS Collaboration}

\bigskip

M.~Aaboud$^\textrm{\scriptsize 34d}$,    
G.~Aad$^\textrm{\scriptsize 99}$,    
B.~Abbott$^\textrm{\scriptsize 125}$,    
O.~Abdinov$^\textrm{\scriptsize 13,*}$,    
B.~Abeloos$^\textrm{\scriptsize 129}$,    
D.K.~Abhayasinghe$^\textrm{\scriptsize 91}$,    
S.H.~Abidi$^\textrm{\scriptsize 164}$,    
O.S.~AbouZeid$^\textrm{\scriptsize 143}$,    
N.L.~Abraham$^\textrm{\scriptsize 153}$,    
H.~Abramowicz$^\textrm{\scriptsize 158}$
% ...
\bigskip
\\

% ...
$^{13}$Institute of Physics, Azerbaijan Academy of Sciences, Baku; Azerbaijan.\\
% ...
$^{34}$$^{(a)}$Facult\'e des Sciences Ain Chock, R\'eseau Universitaire de Physique des Hautes Energies - Universit\'e Hassan II, Casablanca;$^{(b)}$Centre National de l'Energie des Sciences Techniques Nucleaires (CNESTEN), Rabat;$^{(c)}$Facult\'e des Sciences Semlalia, Universit\'e Cadi Ayyad, LPHEA-Marrakech;$^{(d)}$Facult\'e des Sciences, Universit\'e Mohamed Premier and LPTPM, Oujda;$^{(e)}$Facult\'e des sciences, Universit\'e Mohammed V, Rabat; Morocco.\\
% ...
$^{91}$Department of Physics, Royal Holloway University of London, Egham; United Kingdom.\\
% ...
$^{99}$CPPM, Aix-Marseille Universit\'e, CNRS/IN2P3, Marseille; France.\\
% ...
$^{125}$Homer L. Dodge Department of Physics and Astronomy, University of Oklahoma, Norman OK; United States of America.\\
% ...
$^{129}$LAL, Universit\'e Paris-Sud, CNRS/IN2P3, Universit\'e Paris-Saclay, Orsay; France.\\
% ...
$^{143}$Santa Cruz Institute for Particle Physics, University of California Santa Cruz, Santa Cruz CA; United States of America.\\
% ...
$^{153}$Department of Physics and Astronomy, University of Sussex, Brighton; United Kingdom.\\
% ...
$^{158}$Raymond and Beverly Sackler School of Physics and Astronomy, Tel Aviv University, Tel Aviv; Israel.\\
% ...
$^{164}$Department of Physics, University of Toronto, Toronto ON; Canada.\\
% ...
$^{a}$ Also at  Department of Physics, University of Malaya, Kuala Lumpur; Malaysia.\\
$^{b}$ Also at Borough of Manhattan Community College, City University of New York, NY; United States of America.\\
$^{c}$ Also at California State University, East Bay; United States of America.\\
$^{d}$ Also at Centre for High Performance Computing, CSIR Campus, Rosebank, Cape Town; South Africa.\\
$^{e}$ Also at CERN, Geneva; Switzerland.\\
% ...
$^{*}$ Deceased

\end{flushleft}


\end{document}